\documentclass[aps,epsfig,rotate,preprint]{revtex4}
\usepackage{amssymb}
\usepackage{latexsym}
\usepackage{amsfonts}
\usepackage{graphicx}
\usepackage{epsfig}
\usepackage{amsmath}
\usepackage{float}

\begin{document}

\title{Statistical analysis of the overnight and daytime return}
\author{Fengzhong Wang,$^1$ Shwu-Jane Shieh,$^{1,2}$ Shlomo
Havlin,$^{1,3}$ and H. Eugene~Stanley$^1$} \affiliation{$^1$Center for
Polymer Studies and Department of Physics, Boston University, Boston,
MA 02215 USA\\$^2$Department of International Business, National
Cheng-Chi University, Taipei, Taiwan, R.O.C.\\$^3$Minerva Center and
Department of Physics, Bar-Ilan University, Ramat-Gan 52900, Israel}

\date{4 March 2009 ~~~ wshs.tex}

\begin{abstract}

We investigate the two components of the total daily return
(close-to-close), the overnight return (close-to-open) and the daytime
return (open-to-close), as well as the corresponding volatilities of
the 2215 NYSE stocks from 1988 to 2007. The tail distribution of the
volatility, the long-term memory in the sequence, and the
cross-correlation between different returns are analyzed. Our results
suggest that: (i) The two component returns and volatilities have
similar features as that of the total return and volatility. The tail
distribution follows a power law for all volatilities, and long-term
correlations exist in the volatility sequences but not in the return
sequences. (ii) The daytime return contributes more to the total
return. Both the tail distribution and the long-term memory of the
daytime volatility are more similar to that of the total volatility,
compared to the overnight records. In addition, the cross-correlation
between the daytime return and the total return is also stronger. (iii)
The two component returns tend to be anti-correlated. Moreover, we
find that the cross-correlations between the three different returns
(total, overnight, and daytime) are quite stable over the entire
20-year period.

\end{abstract}

\maketitle

\section{Introduction}

Financial markets are of great importance for economics and
econophysics research
\cite{Mandelbrot63,Bunde94,Mantegna00,Kondor99,Bouchaud03,Johnson03,Takayasu97,Ding93,Wood85,Harris86,Yamasaki05,Wang06,Wang08,Weber07,Eisler08,Jung08,Black73,Cox76,George95,Wang02,Liu99}. A
key topic of the market studies is the price dynamics, which could be
measured by the price change (``return'') and its magnitude
(``volatility'')
\cite{Mantegna00,Kondor99,Johnson03,Bouchaud03,Takayasu97,Ding93,Wood85,Harris86,Yamasaki05,Wang06,Wang08,Weber07,Eisler08,Jung08,Black73,Cox76,George95,Wang02,Liu99}. Especially,
the volatility has important practical implications. For example, it
is the key input for option pricing models such as the classic
Black-Scholes model and Cox, Ross, and Rubinstein binomial models
\cite{Black73,Cox76}. Usually financial markets are closed during the
night, and all news or events in the night are reflected in the
opening price of the next trading day. A day (from former day closing
to current day closing) therefore can be decomposed into two sessions,
overnight (from former day closing to current day opening) and daytime
(from current day opening to closing) sessions. The study of the
returns and the volatilities during these two sessions might provide
new insights towards better understanding of the financial
markets. Practically, this study can help traders to improve trading
strategies at the market opening and closing. It also can help
investors to analyze the dually-traded equities \cite{Wang02}.
 
Recently there were some studies on the returns and volatilities over
sub-day sessions. George and Hwang decomposed the daily return of 200
Japanese stocks and analyzed their volatility patterns
\cite{George95}. Wang et. al. studied 15 stocks which are traded in
both Hong Kong and London but in different hours
\cite{Wang02}. However, there is still lack of a comprehensive
analysis of the overnight and daytime price change for a leading
market such as the New York Stock Exchange (NYSE). For the daily and
high-frequency intraday data, returns and volatilities of stock prices
are well studied
\cite{Mantegna00,Kondor99,Johnson03,Bouchaud03,Takayasu97,Ding93,Wood85,Harris86,Yamasaki05,Wang06,Wang08,Weber07,Eisler08,Jung08,Black73,Cox76,George95,Wang02,Liu99}. These
studies show that the return and volatility distribution decay as
power laws, and the correlations in the returns disappear after few
minutes while the correlations in the volatility time series can exist
upto months and even longer
\cite{Mandelbrot63,Lux96,Muller98,Liu99,Plerou07,Clauset07}. It is of
interest to examine whether these features persist also in the two
component returns and volatilities. Obviously one can assume that the
overnight price change behaves statistically different from the
daytime change. What are the differences? Furthermore, the influence
of the overnight price change on the daytime change is also of
interest and should be examined.

In this paper we examine the daily data for all stocks traded in NYSE.
First we study the fundamental features of the time series,
distribution of the records and the correlations in the sequence.
Three types of functions, power law, exponential, and power law with
an exponential cutoff, are tested for the tail of the volatility
distribution. We find that the power law function fits best for most
stocks. Then we analyze the long-term memory of each stock using the
detrended fluctuation analysis (DFA) method
\cite{Peng94,Bunde00,Hu01,Kantelhardt01}, and find that the long-term
correlations persist in the volatilities of both components. We show
that the distribution and the long-term memory of the daytime
volatility is more similar to the total volatility, compared to the
overnight volatility. Further, we study the cross-correlations between
the three types of returns (total, overnight, and daytime). The two
component returns are found to be weakly anti-correlated but both
overnight and daytime return are strongly correlated with the total
return. Interestingly, we find that this behavior is quite stable
during the entire 20-year period.

\section{Data and variables}

We collect the daily opening and closing prices of all securities that
are listed in NYSE on December 31, 2007, in total 2215 stocks
\cite{Data}. The record starts from January 2, 1962, but many stocks
have a much shorter history. We do not include the data before 1987
period for two reasons. First, from 1962 to 1987 there exist only very
little data, about 6.5\% of all the data points for these 2215
stocks. Second and more important, there was a huge market crash on
October 19, 1987 (``Black Monday''), and after that the market was
adapted in a great extent. Thus, to reduce the complexity of market
structure, we only examine the data from 1988 to 2007, in total 20
years. The length of the 2215 stocks ranges from $N=1000$ to $5000$
trading days. Note that many stocks have splits in the 20-year period,
which causes significant change in the price. Therefore, we adjust all
prices according to the historical splits. The 2215 stocks cover all
industrial sectors, a wide range of the stock market capitalization
(from $6 \times 10^6$ to $5 \times 10^{11}$ dollars), and a wide range
of the average daily volume (from $500$ to $2 \times 10^7$ shares a
day).

\begin{figure*}
  \begin{center}
    \includegraphics[width=\textwidth, angle = 0]{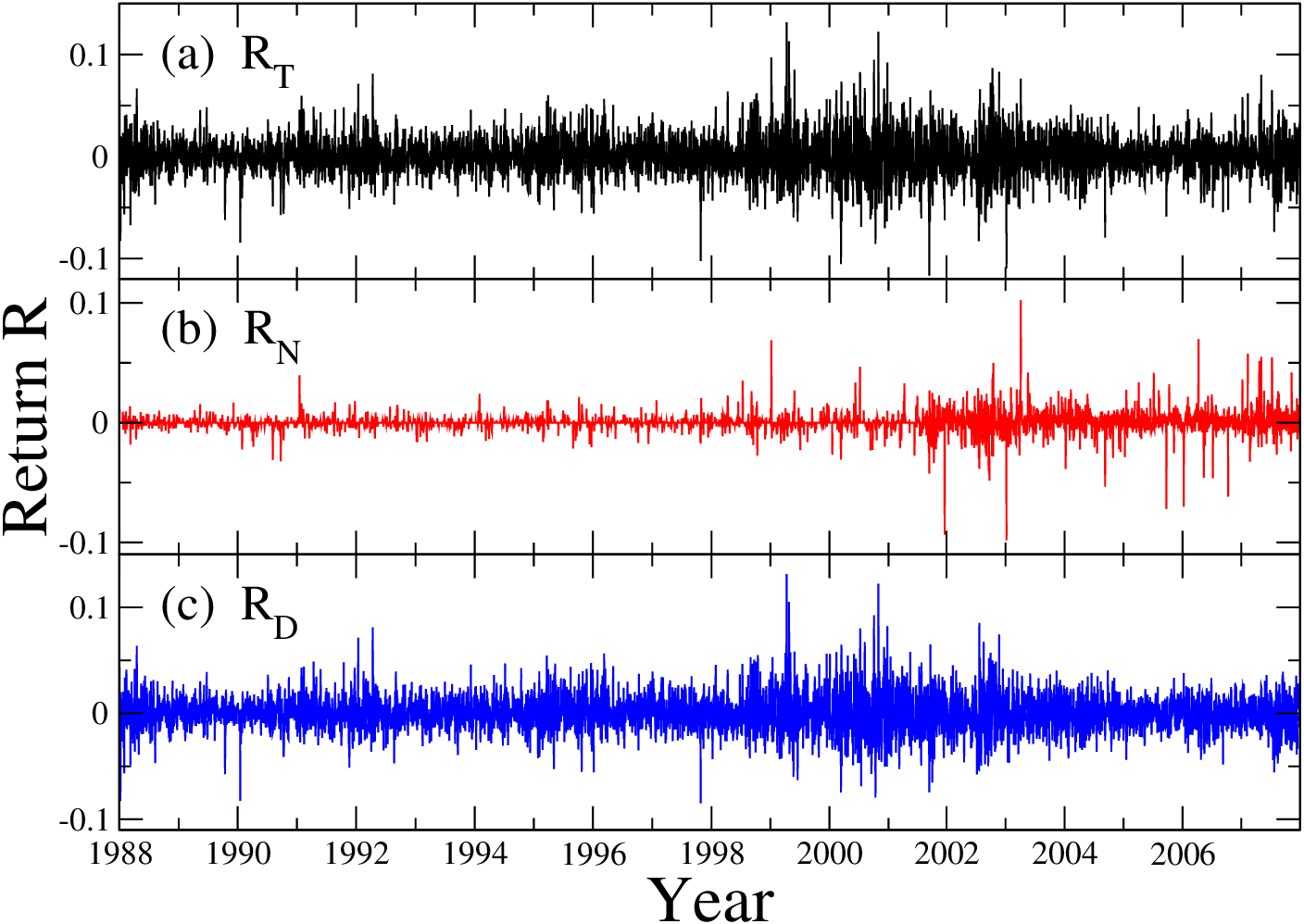}
  \end{center}
  \caption{(Color online) Illustration of the return time
    series. Three types of return, (a) the total return $R_T$, (b) the
    overnight return $R_N$, and (c) the daytime return $R_D$ of a
    typical stock, AA, are shown. We can see that the fluctuations of
    $R_N$ are relatively weaker, and the curve of $R_D$ is more
    similar to that of $R_T$.}
  \label{fig1}
\end{figure*}

Now we define two basic measures, return $R$ and volatility $V$. The
daily return is the logarithmic change of the successive daily
closing prices (``total return''),
\begin{equation}
R_T(t)\equiv \ln(p^{close}(t) / p^{close}(t-1));
\label{total_return.eq}
\end{equation}
the return over the overnight session (``overnight return'') is
\begin{equation}
R_N(t)\equiv \ln(p^{open}(t) / p^{close}(t-1));
\label{overnight_return.eq}
\end{equation}
and the return over the daytime session (``daytime return'') is
\begin{equation}
R_D(t)\equiv \ln(p^{close}(t) / p^{open}(t)).
\label{daytime_return.eq}
\end{equation}
Here $p^{close}(t)$ is the closing price and $p^{open}(t)$ is the
opening price at day $t$. Note that $R_T(t)=R_N(t)+R_D(t)$, $R_D(t)$
and $R_N(t)$ are in the same day and $R_D(t)$ is after
$R_N(t)$. Fig. \ref{fig1} shows the three types of return for a
typical stock AA (Alcoa, Inc.) from 1988 to 2007. The volatility is
defined as the absolute value of the return
\cite{Yamasaki05,Wang06,Wang08}, i.e.
\begin{equation}
V\equiv |R|.
\label{volatility.eq}
\end{equation}
Thus, corresponding to the three types of return, we have three types
of volatility, the total volatility $V_T$, the overnight volatility
$V_N$, and the daytime volatility $V_D$.

\section{Tail of volatility distribution}

The tail distribution accounts for large fluctuations and events which
are very important for risk analysis. By the definition
[Eq. (\ref{volatility.eq})], the volatility aggregates both positive
and negative returns and has better statistics. In addition, the
distribution of the return is approximately symmetric in the two tails
\cite{Liu99}. Therefore we focus on the tail distribution of the
volatility. As a stylized fact of econophysics research, the
cumulative distribution function (CDF) \cite{Note1} of volatilities
has a ``fat tail'' which is usually characterized by a power law
\cite{Mandelbrot63,Lux96,Muller98,Liu99,Plerou07,Clauset07},
\begin{equation}
P(x) \sim x^{-\zeta},
\label{power_law.eq}
\end{equation}
where $\zeta$ is the tail exponent. A classical approach to fit the
tail is using the Maximum Likelihood Estimator, which is called Hill
estimator for a power law tail \cite{Hill75,Plerou07,Clauset07}. The
goodness-of-fit is tested by the Kolmogorov-Smirnov (KS) statistic $D$
\cite{Stephens74,Engle98}, the maximum absolute difference between the
cumulative distribution of the measured distribution $P(x)$ and that
of the fit $S(x)$, i.e.,
\begin{equation}
D\equiv \max(|P(x)-S(x)|),
\label{ks.eq}
\end{equation}
for all volatility values in the tail \cite{Note2}. When $D$ is larger
than a certain value, which is called critical value ($CV$), the null
hypothesis that the distribution follows a power law is rejected. The
$CV$ is determined by the significance level and data size $N$. In
this paper we choose significance level of $1\%$ and the corresponding
$CV=1.63/\sqrt{N}$.

To further test the volatility tail, we also try two other
distribution functions in the same range and using the same
method. One is the exponential distribution function,
\begin{equation}
P(x) \sim e^{-x/x^\ast},
\label{exponential.eq}
\end{equation}
where $x^\ast$ is a characteristic scale. The other is a power law
function with an exponential cutoff,
\begin{equation}
P(x) \sim x^{-\zeta}\cdot e^{-x/x^\ast}.
\label{pl_cutoff.eq}
\end{equation}

\begin{table}
  \caption{Number of good fit of the volatility tail distribution for
  the 2215 NYSE stocks. Good fit refers to the cases where the null
  hypothesis is not ruled out for $1\%$ significance level.}
  \begin{center}
    \begin{tabular*}{0.65\textwidth}{@{\extracolsep{\fill}} | c | c | c | c | }
      \hline
      Volatility V & $V_T$ & $V_N$ & $V_D$ \\
      \hline
      Power law & 2066 & 1868 & 2066 \\
      \hline
      Exponential & 1693 & 644 & 1756 \\
      \hline
      Power law with cutoff & 1755 & 1772 & 1728 \\
      \hline
    \end{tabular*}
  \end{center}
  \label{table1}
\end{table}

We examine the tail distribution of $V_T$, $V_N$ and $V_D$ for the
2215 NYSE stocks. The number of fit that the null hypothesis was valid
under $1\%$ significant level (``good fit'') is listed in Table
\ref{table1}. For the power law distribution, only a small portion
($10\%$) of the three types of volatlities are ruled out, which
manifests that the tail is well characterized by the power law
function for the broad market. For the exponential hypothesis, almost
half ($38\%$) of all cases are ruled out. Moreover, about $98\%$ out
of the good exponential fits, the power law hypothesis is not ruled
out either. As a whole, the exponential function is poor for
characterizing the tail, compared to the power law function. For the
power law with an exponential cutoff, the percentage of good fit is
$79\%$ over the three volatilities, which is slightly lower than that
for the power law. Besides, $99\%$ out of them do not reject the power
law hypothesis either. Therefore, we conclude that the power law is
the best among the three distributions.

\begin{figure*}
  \begin{center}
    \includegraphics[width=\textwidth, angle = 0]{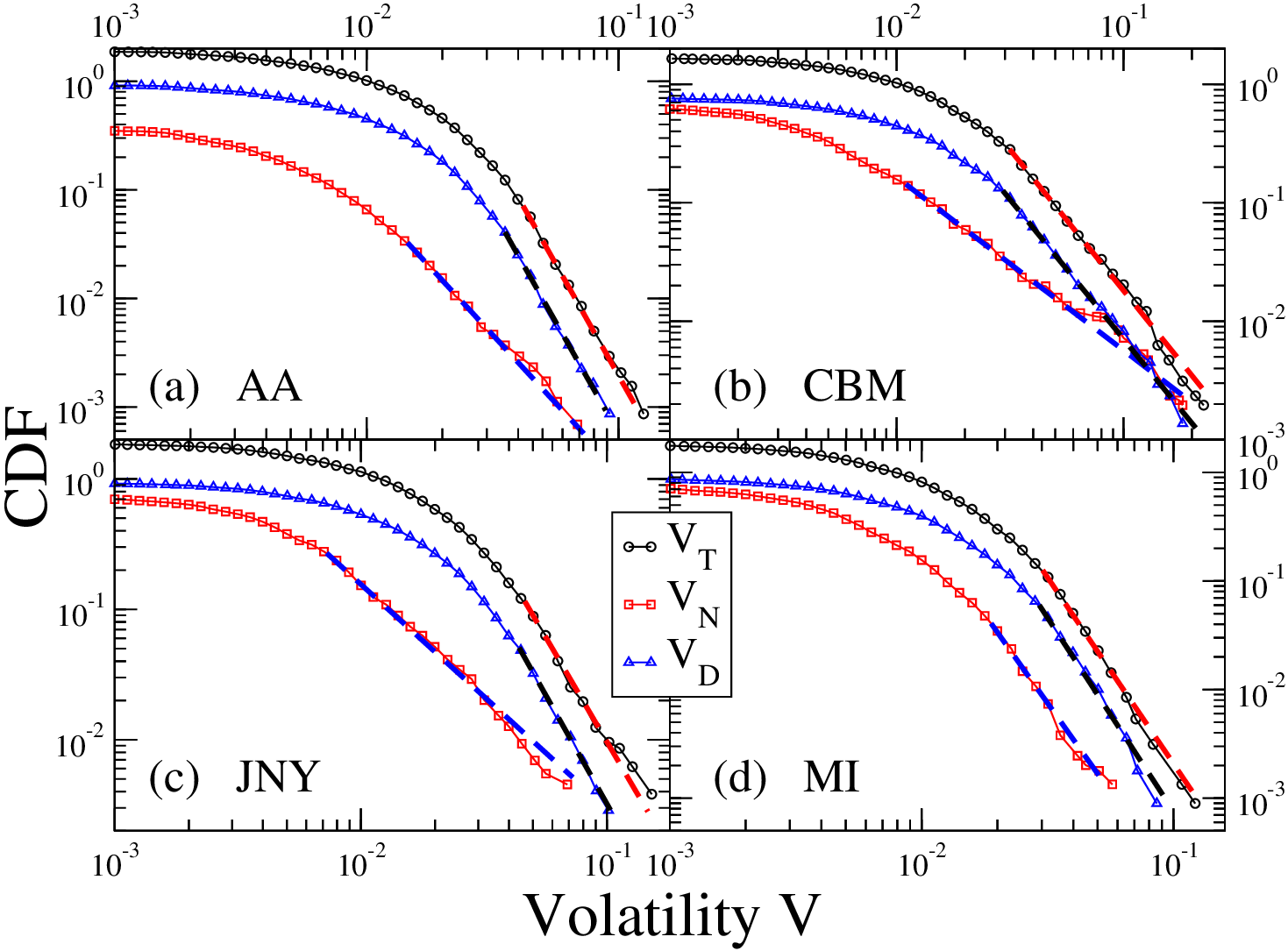}
  \end{center}
  \caption{(Color online) Typical cumulative distribution of the
    volatilities and power law fit to the tails. For four typical
    stocks, (a) AA, (b) CBM, (c) JNY, and (d) MI, three types of
    volatility, total volatility $V_T$ (circles), overnight volatility
    $V_N$ (squares), and daytime volatility $V_D$ (triangles) are
    demonstrated. The dashed lines are power law fits to the
    distribution tails. Note that the curves for $V_T$ (circles)
    almost coincide with those of $V_D$ and thus they are vertically
    shifted for better visibility.}
  \label{fig2}
\end{figure*}

\begin{figure*}
  \begin{center}
    \includegraphics[width=\textwidth, angle = 0]{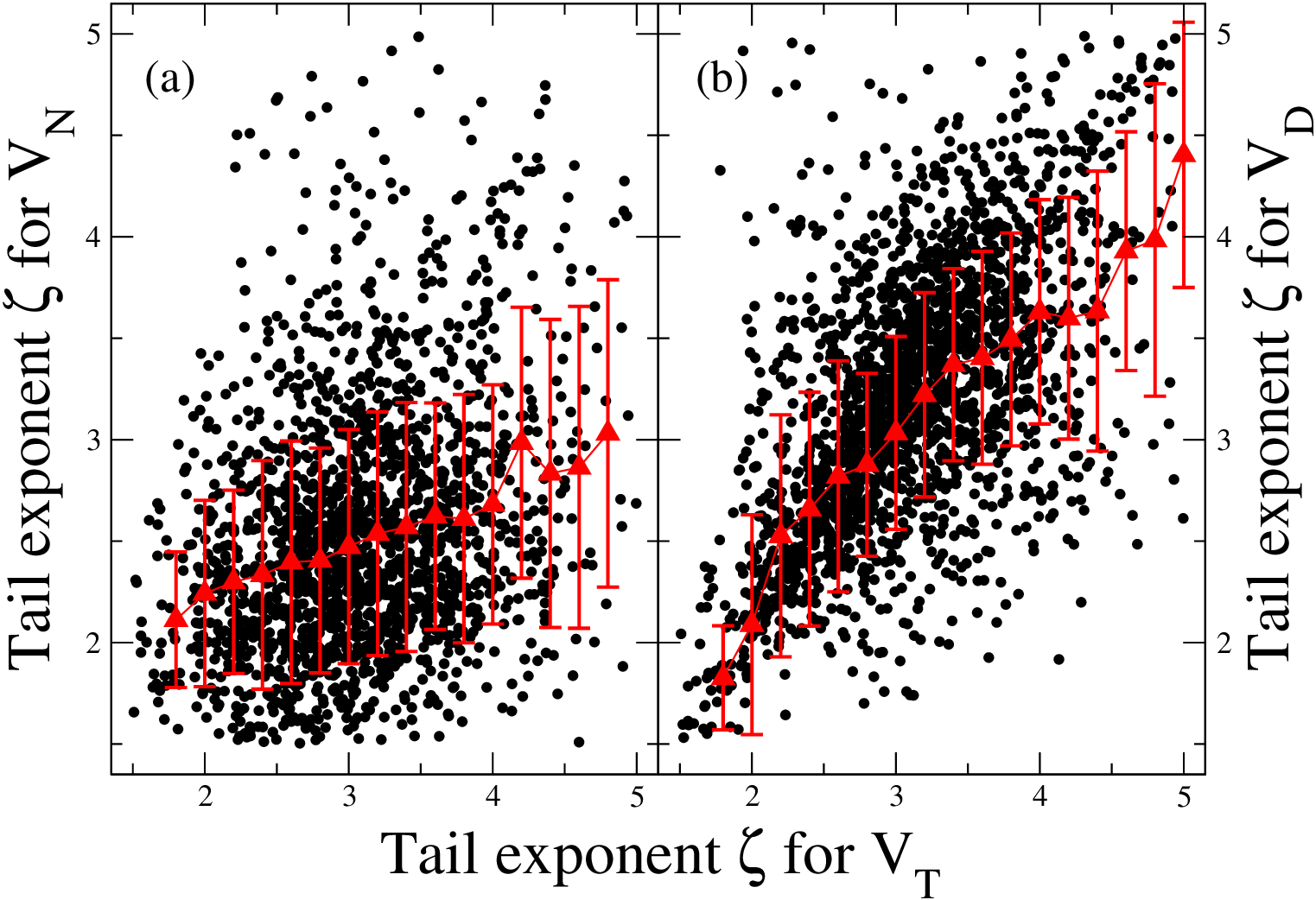}
  \end{center}
  \caption{(Color online) Relation between the tail exponent $\zeta$
    for the total volatility $V_T$ and that for the two component
    volatilities, (a) the overnight volatility $V_N$ and (b) the
    daytime volatility $V_D$. A point represents a stock which has
    good power law fit to the tail for the corresponding two types of
    volatilities. 1812 out of the 2215 NYSE stocks are exhibited in
    panel (a) and 2001 stocks are exhibited in panel (b). To show the
    tendency, we divide the entire data set into equal-width subsets
    according the value of $\zeta$ for $V_T$ and calculate the mean
    values and standard deviations in these subsets, as shown by the
    triangles and the error bars respectively. Both cases clearly show
    tendencies but that for the daytime volatility is stronger,
    indicating $V_T$ is more influenced by $V_D$. Moreover, $\zeta$
    for all three types of volatilities are distributed in a
    relatively narrow range and centered around $3$.}
  \label{fig3}
\end{figure*}

In Fig. \ref{fig2}, we plot the CDF of $V_T$, $V_N$, and $V_D$ for
four typical stocks, namely, Alcoa, Inc. (AA), Cambrex Corp. (CBM),
Jones Apparel Group, Inc. (JNY), and Marshall \& Ilsley
Corp. (MI). These stocks belong to diverse industrial sectors and
their capitalization vary in a wide range, from 27 billion dollars
for AA to 0.25 billion dollars for MI. As seen in Fig. \ref{fig2}, the
tails are well fitted by power laws. Interestingly, the tails of $V_D$
almost always decay faster than the tails of $V_N$, and $V_T$ lies
between the two component volatilities. Moreover, the log-log slope
(tail exponent $\zeta$) of $V_T$ is closer to that of $V_D$,
indicating the daytime return contributes more to the total return. To
test this finding for the broad market, we plot in Fig. \ref{fig3} the
relation between the tail exponent $\zeta$ of $V_T$ and $\zeta$ of the
two component volatilities for the 2215 stocks. Both scatter plots
show certain dependence (as shown by the solid curves, which are
averages over different bins of $\zeta$ of $V_T$), but the correlation
between $V_T$ and $V_D$ is obviously stronger, which is consistent
with Fig. \ref{fig2}. For all three types of volatilities, $\zeta$ is
distributed in a certain range from $1.5$ to $5$, and centered around
$3$. The averages of $\zeta$ are: $\langle \zeta \rangle \approx 2.6$
for $V_N$ is lower than $\langle \zeta \rangle \approx 3.2$ for $V_D$,
while $\langle \zeta \rangle \approx 3.1$ for $V_T$ is between the two
component volatilities and it is slightly smaller than that for
$V_D$. In this paper $\langle ... \rangle$ stands for the average over
the data set. This behavior suggests that the daytime return
influences the total return more than the overnight return.

\section{Correlations in returns and volatilities}

After analyzing the volatility distribution, a question naturally
arises, how these values are organized in the time sequence? For the
investors, the temporal structure is of special interest because it
determines how and when to trade. The time organization in a time
series can be characterized by the two-point correlation. It is known
that the total return has only short-term correlations and the total
volatility has long-term correlations
\cite{Mandelbrot63,Lux96,Muller98,Liu99,Plerou07,Clauset07}. Now we
examine the correlations in each of their two components (overnight
and daytime).

\begin{figure*}
  \begin{center}
    \includegraphics[width=\textwidth, angle = 0]{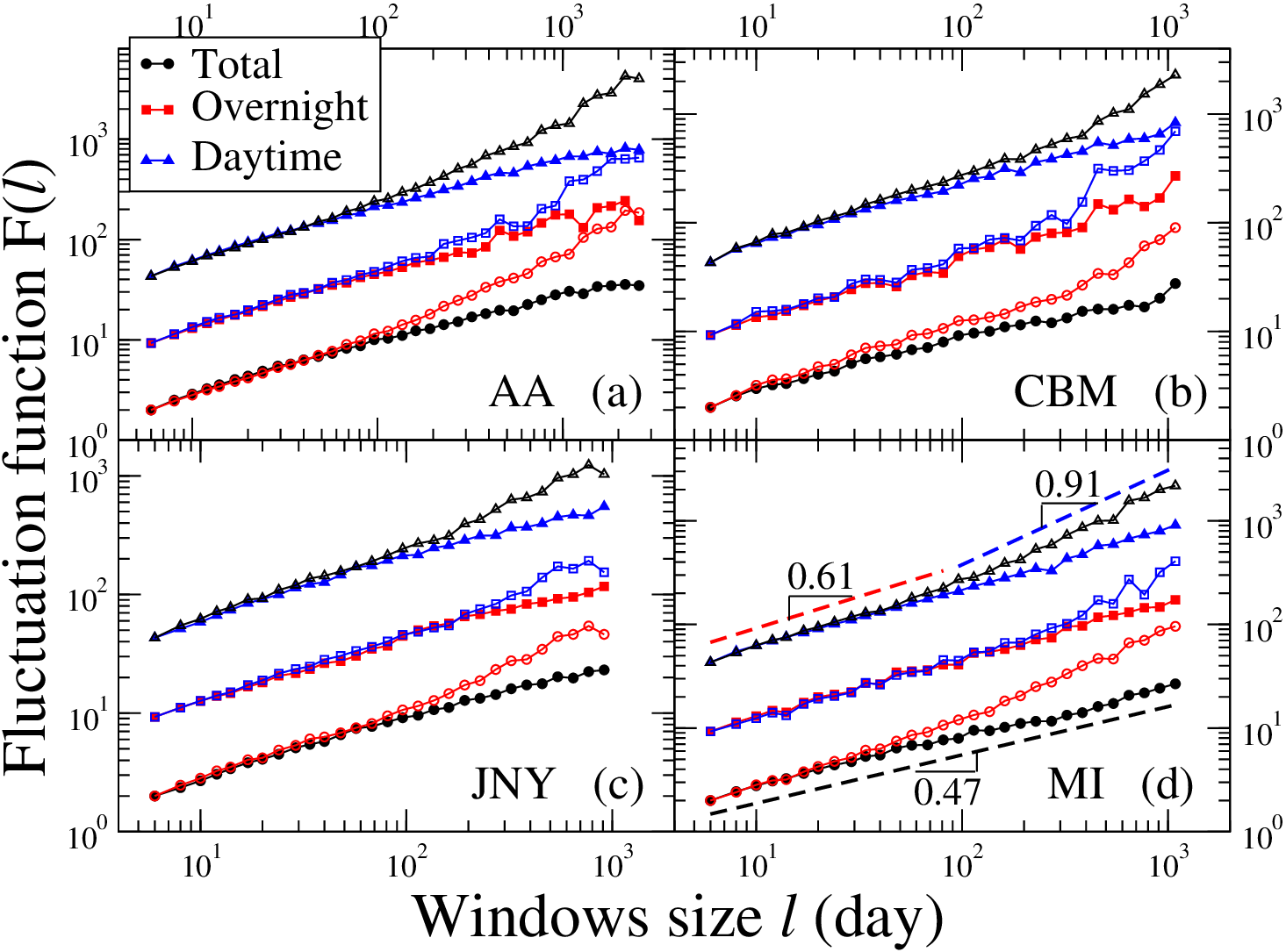}
  \end{center}
  \caption{(Color online) DFA fluctuation function $F$ vs. windows
    size $\ell$ for the returns (filled symbols) and the volatilities
    (open symbols). The four panels are for stocks AA, CBM, JNY, and
    MI respectively. For each case, three types of data, total
    (circles), overnight (squares), and daytime (triangles) are
    shown. Note that the curves are vertically shifted for better
    visibility. To obtain the correlation exponent $\alpha$, we fit
    all curves with power laws, as illustrated by the dashed lines in
    panel (d). For the volatility, the exponent $\alpha$ is
    significantly different for short and long time scales, thus we
    split the entire range into two regimes and fit them separately.}
  \label{fig4}
\end{figure*}

It is well known that financial time series are usually
non-stationary. In such cases, the conventional methods for
correlations such as auto-correlation and spectral analysis have
spurious effects. To avoid the artifact correlations arising from
non-stationarity, we employ the DFA method, which is based on the idea
that a correlated time series could be mapped to a self-similar
process by integration, and removing systematically trends in order to
detect the long-term correlations in the time series
\cite{Peng94,Bunde00,Hu01,Kantelhardt01}. After removing polynomial
trends in every equal-size box of $\ell$ points, DFA computes the
root-mean-square fluctuation $F(\ell)$ of a time series and determine
the correlation exponent $\alpha$ from the scaling function
\begin{equation}
F(\ell)\sim \ell^\alpha,
\label{dfa.eq}
\end{equation} 
where the exponent $\alpha\in(0,1)$, called correlation exponent,
characterizes the auto-correlation in the sequence. It is uncorrelated
if $\alpha=0.5$, positively correlated if $\alpha>0.5$ and
anti-correlated if $\alpha<0.5$. In Figure \ref{fig4}, we plot DFA
curves for the returns and volatilities of the total, overnight, and
daytime sequences for four typical stocks. The values of $\alpha$ are
obtained by the power law fit to the fluctuation function, as
illustrated by the dashed lines in Fig. \ref{fig4}(d). For all three
types of returns, $\alpha$ is close to 0.5 and therefore there are no
long-term correlations. For the volatilities, the fluctuation function
is more complicated. The slopes (in log-log scale) of different
regions are significantly different. Thus, we divide the whole curve
into two equal-size regions in the logarithmic scale and fit them
separately, as shown by the dashed lines in Fig. \ref{fig4}(d).

\begin{figure*}
  \begin{center}
    \includegraphics[width=\textwidth, angle = 0]{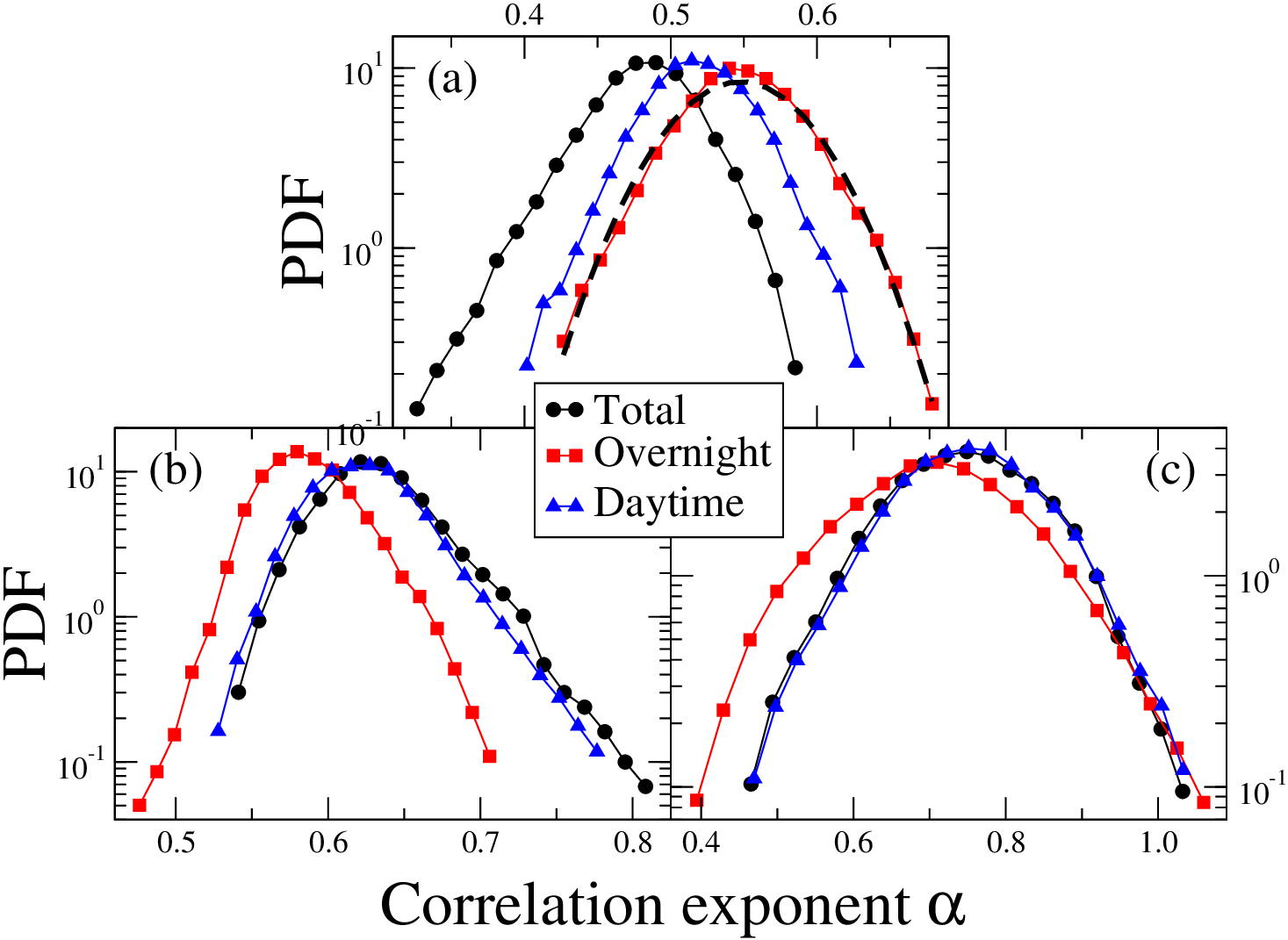}
  \end{center}
  \caption{(Color online) Distribution of the correlation exponent
    $\alpha$ for (a) the returns, (b) the volatilities of the short
    time scales, and (c) the volatilities of the long time
    scales. Three types of returns and volatilities, total (circles),
    overnight (squares), and daytime (triangles) are shown. All
    distributions approximately follow the normal distribution. For
    example, a normal distribution fit on the overnight return is
    shown by the dashed line in panel (a). For the volatilities, the
    curves for the total and the daytime almost collapse into a single
    curve in panels (b) and (c), suggesting that the total return is
    more influenced by the daytime return.}
  \label{fig5}
\end{figure*}

To test the universality of our findings, we plot in Fig. \ref{fig5}
the probability density function (PDF, which is the derivative of CDF)
of $\alpha$ for the three returns as well as for the short and long
time scales of the volatilities. For the returns [Fig. \ref{fig5}(a)],
the distributions are centered around $0.5$, $\alpha=0.48\pm0.04$ for
the total, $\alpha=0.55\pm0.05$ for the overnight and
$\alpha=0.52\pm0.04$ for the daytime. Here and in the following, the
error bars are the standard deviations over all 2215 stocks. These
error bars are quite small representing quite narrow
distributions. This result is consistent with earlier studies, where
no long-term correlations were found for the returns \cite{Liu99}. For
the volatilities at short time scales [Fig. \ref{fig5}(b)], the
distributions are centered around $0.6$, $\alpha=0.63\pm0.04$ for the
total \cite{Liu99}, $\alpha=0.59\pm0.03$ for the overnight and
$\alpha=0.63\pm0.04$ for the daytime. For the volatilities at long
time scales [Fig. \ref{fig5}(c)], $\alpha=0.75\pm0.10$ for the total
\cite{Liu99}, $\alpha=0.71\pm0.12$ for the overnight and
$\alpha=0.75\pm0.10$ for the daytime. For all time scales, the
volatility $\alpha$ values are significantly larger than 0.5,
suggesting long-term correlations in the volatility sequences. In
addition, the $\alpha$ values of the long-term scales are
systematically larger than that of the short-term scales. This
multiscaling behavior indicates that the correlation becomes stronger
for longer times. Moreover, all distributions are relatively narrow
for both returns and volatilities, suggesting a universal feature over
the entire market. We also see that the curves of the total and
daytime almost collapse onto a single curve, while the curve of the
overnight departs away from them, supporting again that the daytime
return contributes more than the overnight return to the total return.

\begin{table}
  \caption{Cross-correlation between the $\alpha$ values of the three
  types of returns and volatilities for the 2215 NYSE stocks. We
  divide the 2215 stocks into 10 equal-size subsets and calculate the
  cross-correlation for every subset. The error bar is the
  corresponding standard deviation of the 10 cross-correlations. The
  value in the parenthesis is the corresponding cross-correlation
  between two shuffled $\alpha$ records.}
  \begin{center}
    \begin{tabular*}{1.00\textwidth}{@{\extracolsep{\fill}} | c | c | c | c | }
      \hline
      Cross-correlation C & C(Total, Overnight) & C(Total, Daytime) & C(Overnight, Daytime) \\
      \hline
      Return & $0.25 \pm 0.07$ & $0.51 \pm 0.08$ & $0.48 \pm 0.08$ \\
      & (-0.00) & (-0.03) & (0.02) \\
      \hline
      Volatility (short time scales) & $0.29 \pm 0.09$ & $0.80 \pm 0.04$ & $0.24 \pm 0.04$ \\
      & (0.04) & (-0.02) & (0.03) \\
      \hline
      Volatility (long time scales) & $0.56 \pm 0.06$ & $0.90 \pm 0.02$ & $0.52 \pm 0.05$ \\
      & (0.01) & (0.02) & (-0.02) \\
      \hline
    \end{tabular*}
  \end{center}
  \label{table2}
\end{table}

Now we address the question if there is a relation between the
correlation exponents $\alpha$ of the two components of the return and
volatility. If a certain stock has large (small) $\alpha$ for one
component, does it have also large (small) $\alpha$ in the other
component or in the total? To test this, we employ the
cross-correlation function to quantitatively compare them. The
cross-correlation (also called the Pearson Coefficient) between
variable $x$ and $y$ is
\begin{equation}
C(x,y) \equiv \frac{\langle x \cdot y \rangle - \langle x \rangle
\cdot \langle y \rangle}{\sigma(x) \cdot \sigma(y)}.
\label{cross_correlation.eq}
\end{equation}
Here $\sigma$ stands for the standard deviation, i.e., for variable
$x$, $\sigma(x) \equiv \sqrt{\langle x^2 \rangle - \langle x
\rangle^2}$. For our case, $x$ and $y$ are vectors representing the
three sequences of $\alpha$ (total, overnight, daytime) for the
return, short time and long time volatilities for all companies. The
companies are in the same order for all sequences. As shown in Table
\ref{table2}, all cross-correlations are significantly larger than
that of shuffled records (values in the parenthesis), suggesting
strong relations between the different returns or volatilities. Note
again that the total-daytime pair is always the strongest one, which
is in agreement with the assumption that the total return and
volatility are significantly more influenced by the daytime return and
volatility, than by the overnight return and volatility.

\section{Relation between total, overnight and daytime returns}

\begin{figure*}
  \begin{center}
    \includegraphics[width=\textwidth, angle = 0]{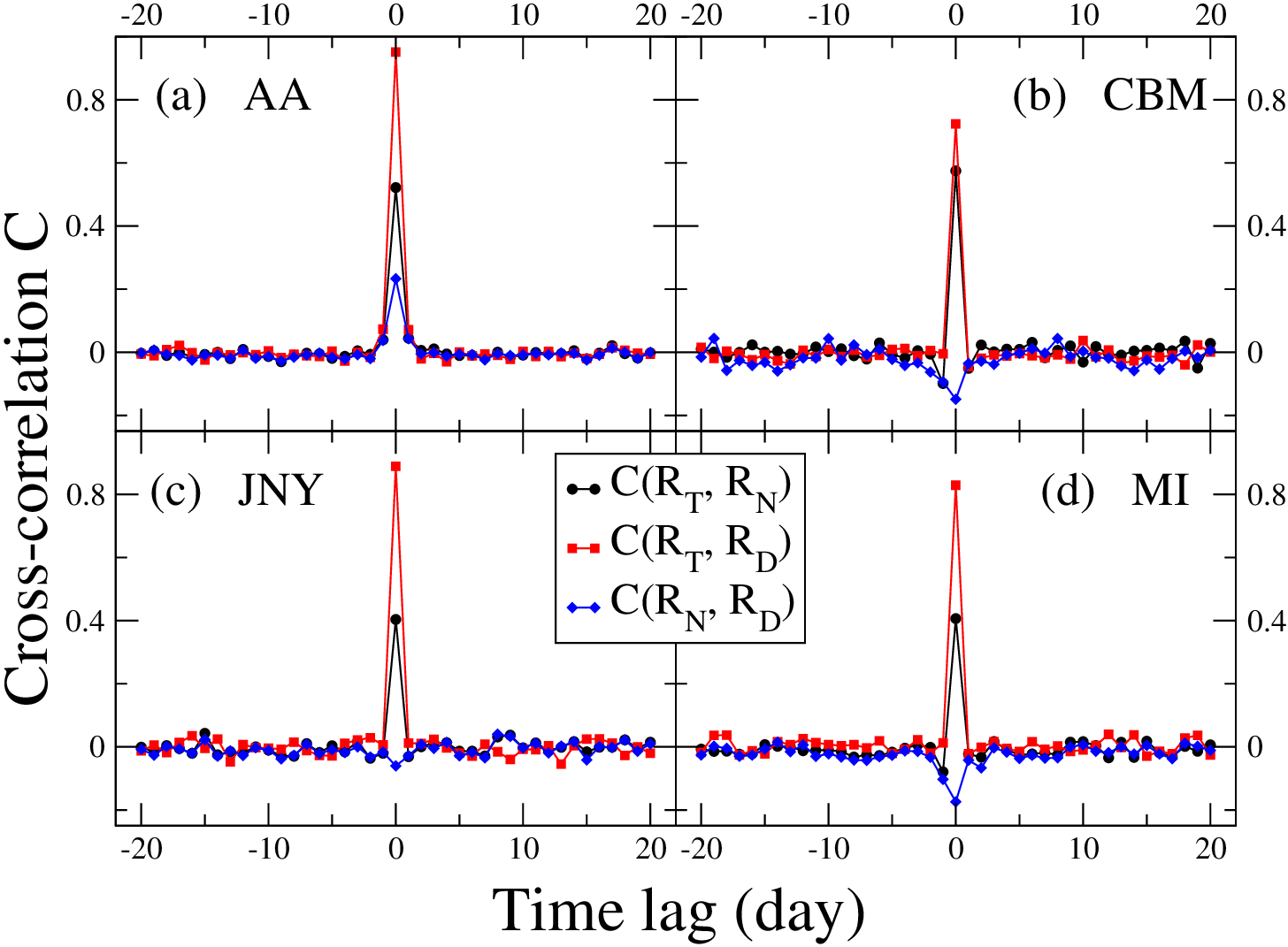}
  \end{center}
  \caption{(Color online) Reliability of the cross-correlations
    between the three types of returns, the total return $R_T$, the
    overnight return $R_N$, and the daytime return $R_D$. For all four
    stocks (a) AA, (b) CBM, (c) JNY, and (d) MI, the two
    cross-correlations with respect to the total return are
    significant larger than their cross-correlations with the time
    lags, which suggests both component returns are strongly
    positively correlated to the total return. However, the
    cross-correlation between the two component returns varies with
    the stocks, e.g., it is positive for AA and negative for MI.}
  \label{fig6}
\end{figure*}

The overnight return and the daytime return are the price changes over
different sessions of a trading day, and they make the total
return. It is interesting to examine now if the three returns of the
same stock are cross-correlated. This will test the question, e.g.,
how changes in the day time are related to those of night time or the
total. The cross-correlation function
[Eq. (\ref{cross_correlation.eq})] examines the two time series
without any time lag. However, there might be some time delays between
two time series, and therefore we shift the two sequences by time lag
$\Delta t$ to test this possibility. Moreover, the comparison between
the cross-correlations with different lags allows us to examine the
significance of a cross-correlation value. Therefore, we use the
generalized cross-correlation with the time lag $\Delta t$, i.e.,
\begin{equation}
C_{\Delta t}(x,y)\equiv \frac{\langle x(t)\cdot y(t+\Delta t) \rangle
-\langle x \rangle \cdot \langle y \rangle}{\sigma(x) \cdot \sigma(y)}
\label{cross_correlation_lag.eq}
\end{equation}
 between two time series $x(t)$ and $y(t)$. Note that
Eq. (\ref{cross_correlation.eq}) is the special case of
Eq. (\ref{cross_correlation_lag.eq}) with $\Delta t=0$. In general one
tests the position of the maximum (minimum if it is anti-correlated)
of $C_{\Delta t}$ which may occur at $\Delta t=\tau$ and $\tau$ is
called the time delay \cite{Yamasaki08}. Here we find that the maximum
of $C_{\Delta t}$ is always for $\Delta t=0$ (as shown in Figure
\ref{fig6}).

In this paper we use $C_{\Delta t}$ to test the significance of the
cross-correlation at $\Delta t=0$. If $C_{\Delta t=0}$ is
significantly different (higher or lower) from $C_{\Delta t\neq0}$,
the cross-correlation can be regarded as reliable. Quantitatively, we
use the standard deviation of $C_{\Delta t\neq0}$ values over the
range $-20\leq \Delta t \leq20$, $\sigma(C_{\Delta t\neq0})$, to test
the reliability of the cross-correlation \cite{Yamasaki08}. As
examples, we plot in Fig. \ref{fig6} the cross-correlations of three
pairs of returns for the four typical stocks, AA, CBM, JNY, and MI
(other stocks have similar features). For both $C(R_T, R_N)$ and
$C(R_T, R_D)$, the cross-correlations at $\Delta t=0$ are more than 10
times higher than their $\sigma(C_{\Delta t\neq0})$ so they are very
robust. However, for $C(R_N, R_D)$, the cross-correlations vary with
the stock. Some of them have significant cross-correlation values but
some of them are in the range of their $\sigma(C_{\Delta
t\neq0})$. Since $R_N$ and $R_D$ covers different periods, there could
be some strong correlations or almost independent, it is reasonable
that the cross-correlation varies in a wide range. On the other hand,
$R_T$ always shares part of changes with its two component returns and
deduce strong positive cross-correlations.

\begin{figure*}
  \begin{center}
    \includegraphics[width=\textwidth, angle = 0]{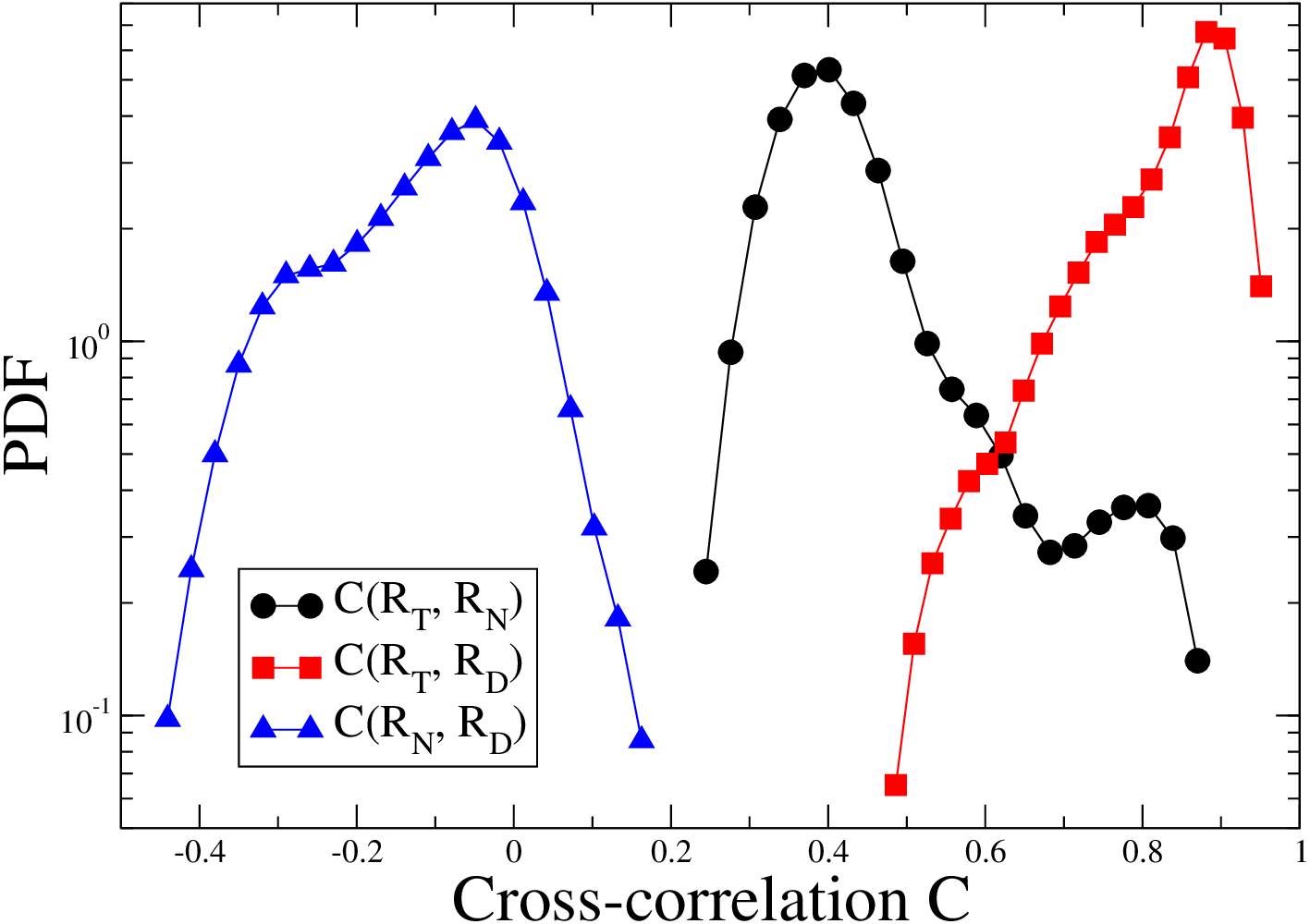}
  \end{center}
  \caption{(Color online) Distribution of the cross-correlations $C$
    between the returns $R_T$, $R_N$, and $R_D$ for the 2215 NYSE
    stocks. Both cross-correlations with respect to the total return
    are significantly larger than $0$ and that for the daytime return
    is stronger, suggesting that the total return is more correlated
    to the daytime return. The cross-correlation between the two
    component returns is relatively more distributed towards negative
    values, indicating the two component returns tend to be
    anti-correlated.}
  \label{fig7}
\end{figure*}

Next we examine the three pairs of cross-correlations $C_{\Delta t=0}$
for all the 2215 stocks (in the following, the function $C$ refers to
$C_{\Delta t=0}$ if the $\Delta t$ subscript is missing). Their
distributions are plotted in Fig. \ref{fig7}. For each pair, the
cross-correlations are distributed in a certain range. The
cross-correlation between the total return and the daytime return,
$C(R_T, R_D)=0.8\pm0.1$ (mean value and standard deviation over the
2215 stocks), is always the largest value in the three pairs. The
cross-correlation between the total and the overnight, $C(R_T,
R_N)=0.4\pm0.1$, is a still high but significantly smaller than
$C(R_T, R_D)$ values. The cross-correlation between day and night,
$C(R_N, R_D)=-0.1\pm0.1$, is distributed around $0$ with more tendency
to have negative values. In summary, the total return is more
synchronized with the daytime return. It is also interesting to note
that there are significantly more stocks that have negative
correlations between $R_N$ and $R_D$. For example, $567$ out of the
2215 stocks have values of $C(R_N, R_D)<-0.2$. This implies that the
probability is relatively high for a large positive overnight return
to be followed by a large negative daytime return. The overnight
return and the daytime return tend to be slightly anti-correlated, and
the total return usually moves in the same direction as the daytime
return.

\begin{figure*}
  \begin{center}
    \includegraphics[width=\textwidth, angle = 0]{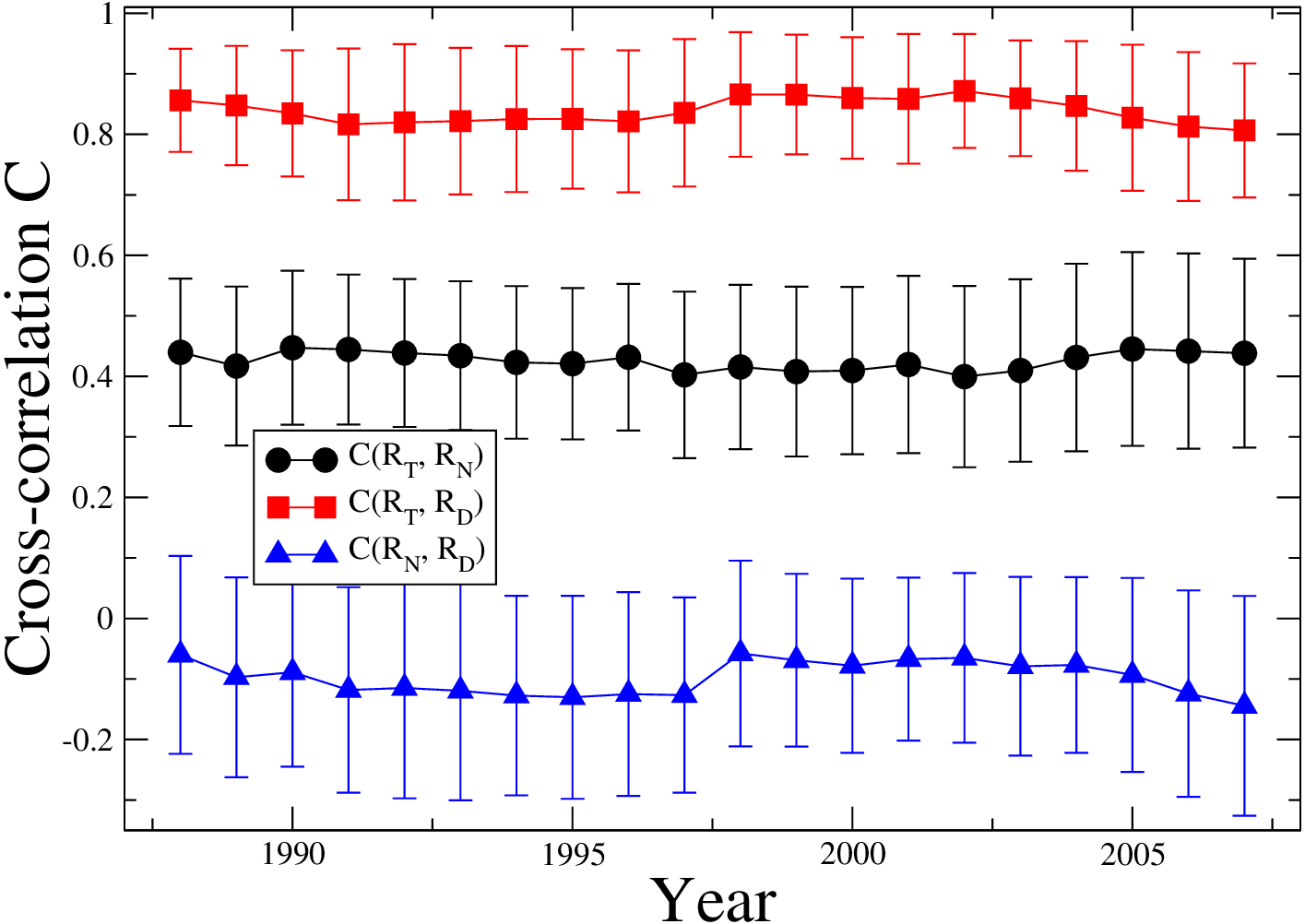}
  \end{center}
  \caption{(Color online) Evolution of the cross-correlations between
    the three returns, $R_T$, $R_N$, and $R_D$, from 1988 to
    2007. Here a point represents the average over the
    cross-correlations of the 2215 NYSE stocks in a 1-year period, and
    the error bar is the corresponding standard deviation. Clearly,
    there are no significant changes for the cross-correlations over
    the 20 years studied.}
  \label{fig8}
\end{figure*}

Due to many factors, such as changes in the regulations or new
technologies, the markets evolves with time. An interesting question
arises, is the cross-correlation stable in the sample years
studied. To test the stability of the cross-correlations, we
recalculate the cross-correlations year by year. The records in 1 year
are enough to calculate the cross-correlation and more importantly,
the equity market in such a short period can be assumed stable. In
Fig. \ref{fig8} we plot the averages and standard deviations (as error
bars) of the cross-correlations over the 2215 stocks against the
year. For the three types of cross-correlations, the curves only
slightly vary with the years and all changes lie within the error
bars. Moreover, the error bars are almost the same for all years,
which clearly shows that the cross-correlation is quite stable over
the 20 years period studied.

\section{Discussion}

Returns and volatilities might be affected by some factors, such as
the market capitalization and the mean volume \cite{Wang08}. To test
this for the entire stock market, we investigate the relation between
the factors, including the capitalization and mean volume, and the
measures, such as the tail exponent $\zeta$, the correlation exponent
$\alpha$, and the cross-correlations between the three returns. There
are some tendencies between these factors and measures. However, most
of these tendencies are in the range of the error bars, which suggests
no significant dependence between the two factors and three
measurements. The behavior of the three measures is quite universal
over the entire market. To better understand the complexity of the
equity market, the connection between different measurements and
factors of stocks might need to be further analyzed.

In summary, we examined the distributions of the total, overnight and
daytime volatility. Compared to the exponential and power law with
cutoff, power law distribution is found to be mostly better. The tail
exponent $\zeta$ is distributed among the different stocks between
$1.5$ and $5$ for the three types of volatility. We also analyzed the
correlations in returns and volatilities of the components, using the
DFA method. For both returns, there are no long-term
correlations. However, for both volatilities, there are long-term
correlations in all time scales and the correlations are even stronger
in the long time scales. For the tail distribution and for the
long-term correlations, the results of the two component returns and
volatilities are similar to the total return and volatility. Moreover,
the records of the daytime are more similar to the total of the same
stock, suggesting that the daytime return contributes more to the
total return. To better compare these similarities, we studied also
the cross-correlations between the different types of return and found
consistent behaviors, i.e., the daytime is more correlated to the
total compared to the night time. Further, the cross-correlation
between the overnight return and the daytime return varies for
different stocks, and interestingly, a significant fraction of the
2215 stocks is far below $0$. This finding suggests that the daytime
return has a considerable probability to strongly anti-correlate with
the overnight return. Furthermore, we examined the cross-correlations
year by year and found that the behavior is quite stable over the
20-year period.

\section*{Acknowledgments}

We thank K. Yamasaki for helpful discussions, the National Science
Foundation and Merck Foundation for financial support, and Shwu-Jane
Shieh thanks the NSF of Taiwan for financial support.


\newpage

\begin{thebibliography}{99}

\bibitem{Bunde94} {\it Fractals in Science}, edited by A. Bunde and
S. Havlin (Springer, Heidelberg, 1994).

\bibitem{Mantegna00} R. Mantegna and H. E. Stanley, {\it Introduction
to Econophysics: Correlations and Complexity in Finance} (Cambridge
University Press, Cambridge, England, 2000).

\bibitem{Kondor99} {\it Econophysics: An Emerging Science}, edited by
I. Kondor and J. Kert\'{e}sz (Kluwer, Dordrecht, 1999).

\bibitem{Bouchaud03} J.-P. Bouchaud and M. Potters, {\it Theory of
Financial Risk and Derivative Pricing: From Statistical Physics to
Risk Management\/} (Cambridge University Press, Cambridge, England,
2003).

\bibitem{Johnson03} N. F. Johnson, P. Jefferies, and P. M. Hui, {\it
Financial Market Complexity} (Oxford University Press, New York,
2003).

\bibitem{Takayasu97}  H. Takayasu, H. Miura, T. Hirabayashi, and
K. Hamada, Physica A {\bf 184}, 127 (1992); H. Takayasu, A. H. Sato,
and M. Takayasu, Phys. Rev. Lett. {\bf 79}, 966 (1997); H. Takayasu
and K. Okuyama, Fractals {\bf 6}, 67 (1998).

\bibitem{Ding93} Z. Ding, C. W. J. Granger, and R. F. Engle, J.
Empirical Finance {\bf 1}, 83 (1993).

\bibitem{Wood85} R. A. Wood, T. H. McInish, and J. K. Ord, J. Financ.
{\bf 40}, 723 (1985).

\bibitem{Harris86} L. Harris, J. Financ. Econ. {\bf 16}, 99 (1986).

\bibitem{Yamasaki05} K. Yamasaki, L. Muchnik, S. Havlin, A. Bunde, and
H. E. Stanley, Proc. Natl. Acad. Sci. U.S.A. {\bf 102}, 9424 (2005).

\bibitem{Wang06} F. Wang, K. Yamasaki, S. Havlin, and H. E. Stanley,
Phys. Rev. E {\bf 73}, 026117 (2006); F. Wang, P. Weber, K. Yamasaki,
S. Havlin, and H. E. Stanley, Eur. Phys. J. B {\bf 55}, 123 (2007);
W.-S. Jung, F. Z. Wang, S. Havlin, T. Kaizoji, H.-T. Moon and
H. E. Stanley, Eur. Phys. J. B {\bf 62}, 113 (2008);
I. Vodenska-Chitkushev, F. Z. Wang, P. Weber, K. Yamasaki, S. Havlin,
and H. E. Stanley, Eur. Phys. J. B {\bf 61}, 217 (2008).

\bibitem{Wang08} F. Wang, K. Yamasaki, S. Havlin and H. E. Stanley,
Phys. Rev. E {\bf 77}, 016109 (2008); F. Wang, K. Yamasaki, S. Havlin
and H. E. Stanley, Phys. Rev. E {\bf 79}, 016103 (2009).

\bibitem{Weber07} P. Weber, F. Wang, I. Vodenska-Chitkushev,
S. Havlin, and H. E. Stanley, Phys. Rev. E 76, 016109 (2007).

\bibitem{Eisler08} Z. Eisler, I. Bartos and J. Kert\'{e}sz,
Adv. Phys. {\bf 57}, 89 (2008).

\bibitem{Jung08} W.-S. Jung, O. Kwon, F. Wang, T. Kaizoji, H.-T. Moon,
and H. E. Stanley, Physica A {\bf 387}, 537 (2008).

\bibitem{Black73} F. Black and M. Scholes, J. Polit. Econ. {\bf 81},
637 (1973).

\bibitem{Cox76} J. C. Cox and S. A. Ross, J. Financ. Econ. {\bf 3},
145 (1976); J. C. Cox, S. A. Ross, and M. Rubinstein, J. Financ. Econ.
{\bf 7}, 229 (1979).

\bibitem{George95} T. J. George and C. H. Hwang,
J. Finan. Quant. Anal. {\bf 30}, 313 (1995).

\bibitem{Wang02} S. S. Wang, O. M. Rui, M. Firth, J. Int. Money
Financ. {\bf 21}, 265 (2002).

\bibitem{Liu99} Y. Liu, P. Gopikrishnan, P. Cizeau, M. Meyer,
C.-K. Peng, and H. E. Stanley, Phys. Rev. E {\bf 60}, 1390 (1999);
V. Plerou, P. Gopikrishnan, L. A. N. Amaral, M. Meyer, and
H. E. Stanley, Phys. Rev. E {\bf 60}, 6519 (1999); V. Plerou,
P. Gopikrishnan, X. Gabaix, L. A. Nunes Amaral, and H. E. Stanley,
Quant. Finance {\bf 1}, 262 (2001); V. Plerou, P. Gopikrishnan, and
H. E. Stanley, Phys. Rev. E {\bf 71 }, 046131 (2005).

\bibitem{Mandelbrot63} B. B. Mandelbrot, J. Business {\bf 36}, 394
(1963).

\bibitem{Lux96} T. Lux, Appl. Finan. Econ. {\bf 6}, 463 (1996).

\bibitem{Muller98} U. A. Muller, M. M. Dacorogna, and O. V. Pictet,
``Heavy Tails in High-Frequency Financial Data,'' in {\it A Practical
Guide to Heavy Tails}, edited by R. J. Adler, R. E. Feldman, and
M. S. Taqqu (Birkh\"{a}user Publishers, 1998), p. 83.

\bibitem{Plerou07} V. Plerou and H. E. Stanley, Phys. Rev. E 76,
046109 (2007).

\bibitem{Clauset07} A. Clauset, C. R. Shalizi, and M. E. J. Newman,
http://arxiv.org/abs/0706.1062v1.

\bibitem{Hill75} B. M. Hill, Ann. Stat. {\bf 3}, 1163 (1975).

\bibitem{Peng94} C.-K. Peng, S. V. Buldyrev, S. Havlin, M. Simons,
H. E. Stanley, and A. L. Goldberger, Phys.  Rev. E {\bf 49}, 1685
(1994).

\bibitem{Bunde00} A. Bunde, S. Havlin, J. W. Kantelhardt, T. Penzel,
J.-H.  Peter, and K. Voigt, Phys. Rev. Lett. {\bf 85}, 3736 (2000).

\bibitem{Hu01} K. Hu, P. Ch. Ivanov, Z. Chen, P. Carpena, and
H. E. Stanley, Phys. Rev. E {\bf 64}, 011114 (2001).

\bibitem{Kantelhardt01} J. W. Kantelhardt, E. Koscielny-Bunde,
H. H. A. Rego, S. Havlin, and A. Bunde, Physica A {\bf 295}, 441
(2001).

\bibitem{Data} Historical stock data is available at
http://finance.yahoo.com. To obtain good statistics, we only choose
the stocks with more than 1000 records.

\bibitem{Note1} In this paper the cumulative distribution function is
actually the complementary cumulative distribution function, $P(x)
\equiv P(x^\prime > x)$, and $P$ is the probability of variable
$x$. For simplicity we call it ``cumulative distribution'' or CDF.

\bibitem{Stephens74} M. A. Stephens, J. Am. Stat. Assoc. {\bf 69}, 730
(1974).

\bibitem{Engle98} R. Engle and J. Russel, Econometrica {\bf 66}, 1127
(1998).

\bibitem{Note2} The range of fit for the Hill estimator is not fixed
\cite{Plerou07,Clauset07}. Here we shift the lower bound of the range,
examine the corresponding KS statistics, and choose one that has the
minimum $D$ value, as suggested by Ref. \cite{Clauset07}. To obtain
good statistics, we use the highest $3\%$ of all data points as the
minimum range. For the three types of volatilities of the 2215 stocks,
the average number of data points in the tail is about $430$ (the tail
averagely covers the highest $10\%$ data points). In order to put all
fits on the same footing, we use the same range for a stock in the two
other types of fit, the exponential and the power law with exponential
cutoff.

\bibitem{Yamasaki08} K. Yamasaki, A. Gozolchiani, and S. Havlin,
Phys. Rev. Lett. {\bf 100}, 228501 (2008).

\end{thebibliography}
\end{document}